\documentclass[
aps,%
12pt,%
final,%
notitlepage,%
oneside,%
onecolumn,%
nobibnotes,%
nofootinbib,%
superscriptaddress,%
noshowpacs]%
{revtex4}%
\usepackage{graphics}
\usepackage[english]{babel}
\usepackage{amssymb,amsmath}

\begin{document}
\title{Associative production of $\Upsilon$ and open charm at LHC.}
\author{\firstname{A.~V.}~\surname{Berezhnoy}}
\email{Alexander.Berezhnoy@cern.ch}
\affiliation{SINP of Moscow State University, Russia}

\author{\firstname{A.~K.}~\surname{Likhoded}}
\email{Anatolii.Likhoded@ihep.ru}
\affiliation{Institute for High Energy Physics, Protvino, Russia}

\begin{abstract}
The yield of  $\Upsilon$ associated with open charm has been estimated with different approaches. The crucial differences between SPS and DPS predictions are discussed.

\end{abstract}

\maketitle
\section{introduction}

Last years the impressive volume of experimental  data on  multiple heavy quarks production were obtained by the experiments at LHC. The LHCb experiment has published studies of  the  $B_c$-meson production~\cite{Aaij:2014ija}, of the  double $J/\psi$-meson production~\cite{Aaij:2011yc}, of the double open charm  production, as well as the $J/\psi$ production associated with open charm~\cite{Aaij:2012dz}.   The LHCb  study of $B_c$ meson is completed by ATLAS, where the candidate to $B_c(2S)$ state has been observed~\cite{Aad:2014laa}.   Also very interesting results on the double $J/\psi$ production have been obtained by the CMS Collaboration~\cite{CMS:2013pph}. 

It is worth, to note that the  double $J/\psi$ production at LHCb can be satisfactorily  described within the standard NRQCD approach~\cite{Berezhnoy:2011xy}, as well  
within  the $k_T$ factorization approach~\cite{Baranov:2011zz}. Contrary to this,  for  other processes of multiple heavy quark production  NRQCD underestimates the  cross section value by the order of magnitude. This could mean,  that in addition  to the single parton scattering (SPS), the double parton scattering  (DPS) should be taken into account. According DPS, approach heavy 	quark pairs are produced independently in different partonic interactions inside the same colliding pair of protons. The simplest variant of this model lead to the following formula for the cross section:
\begin{equation}
\sigma_{A_1 A_2}^\mathrm{DPS}=\frac{1}{m}
\frac{\sigma_{A_1}^\mathrm{SPS}\sigma_{A_2}^\mathrm{SPS}}{\sigma_{eff}},
\label{eq:DPS}
\end{equation}
where $\sigma_{A_1}^\mathrm{SPS}$ and $\sigma_{A_2}^\mathrm{SPS}$ 
 are the cross section values of the processes $A_1$ and $A_2$ within SPS, $m=1$ for different $A_1$ and $A_2$, $m=1/2$ for identical  $A_1$ and $A_2$, and $\sigma_{eff} $ is the parameter of DPS model obtained from the experimental data~\cite{Abe:1997xk,Abazov:2009gc}.
  
The  formula~(\ref{eq:DPS}) is surprisingly successful in  predicting of the cross section values  for the kinematical condition of the LHCb experiment. However,  DPS fails in describing of some differential distributions, and therefore in spite of the fact that DPS is recently  the most successful model in the discussed field, the  problem still exists~\cite{Aaij:2012dz}.

Going back to the double $J/\psi$ production at LHCb it worth to mention,that together with SPS, DPS also could contribute to this process, because    the predictions within SPS and DPS for this case have the same order of magnitude~\cite{Kom:2011bd,Baranov:2011ch,Novoselov:2011ff}.

As a rule it is assumed within the DPS, that both  heavy quark production  and hadronization processes occur independently for the processes $A_1$ and $A_2$. But in principle one could assume
that the soft processes  of hardronization  occur  
mutually. In the last  case the heavy quark and the heavy antiquark from different pair could join to the heavy meson due to the soft interaction followed by the hard production process (see, for example,~\cite{Harland-Lang:2014efa}, where such contribution has been studied for the central exclusive production of $J/\psi$~pair).  Therefore it would be very interesting to  compare experimentally the process of $J/\psi + c$ production, where quarks in  $J/\psi$ can hypothetically origin from different pairs and the process of  $\Upsilon +c$ production.

In this paper we discuss some theoretical aspects of the hadronic 
  $\Upsilon +c$ production, as well as  its observation prospects.

\section{ $\Upsilon +c$ production within SPS}

\begin{figure}
 \centering
\resizebox*{0.7\textwidth}{!}{\includegraphics{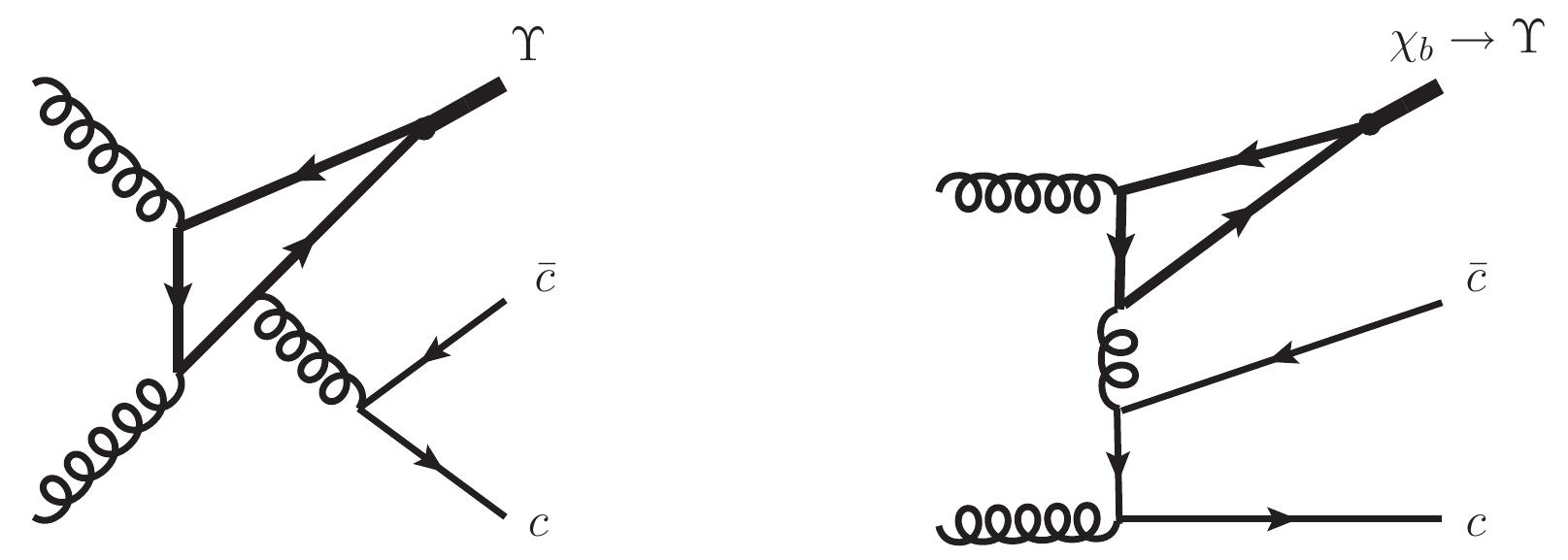}}
\caption{ The examples of  LO diagrams for the $\Upsilon +c$ production in the gluonic fusion.}
\label{fig:ups_cc_diagr}
\end{figure}

There are  6 LO diagrams, which contribute to the direct gluonic production of the associated $\Upsilon$ and $c$  in SPS (see fig.~\ref{fig:ups_cc_diagr}):
\begin{equation}
gg \to \Upsilon + c \bar c.
\end{equation}

Also 10 diagrams contribute to the indirect production of $\Upsilon$ mesons in  $\chi_b$ decays:
\begin{equation}
gg \to \chi_b + c \bar c, \; \chi_b \to \Upsilon.
\label{eq:gg_to_chib_cc}
\end{equation}

For rough estimations it is not necessary to take into account the feed-down from the subprocess~(\ref{eq:gg_to_chib_cc}). Indeed, we expect, that within NRQCD $\sigma(\chi_b+c)/\sigma(\Upsilon+c) \sim 10\% \div 20\%$. Taking into account that $Br(\chi_{b0}\to \Upsilon)\approx 1.8 \%$, $Br(\chi_{b1}\to \Upsilon)\approx 34 \%$ and  $Br(\chi_{b2}\to \Upsilon)\approx 19 \%$,  we obtain that
\begin{equation}
\frac{\sigma(gg \to \chi_b + c \bar c, \; \chi_b \to \Upsilon)}{\sigma(gg \to \Upsilon + c \bar c)} \lesssim 6\%.
\end{equation}

\begin{figure}
 \centering
\resizebox*{0.7\textwidth}{!}{\includegraphics{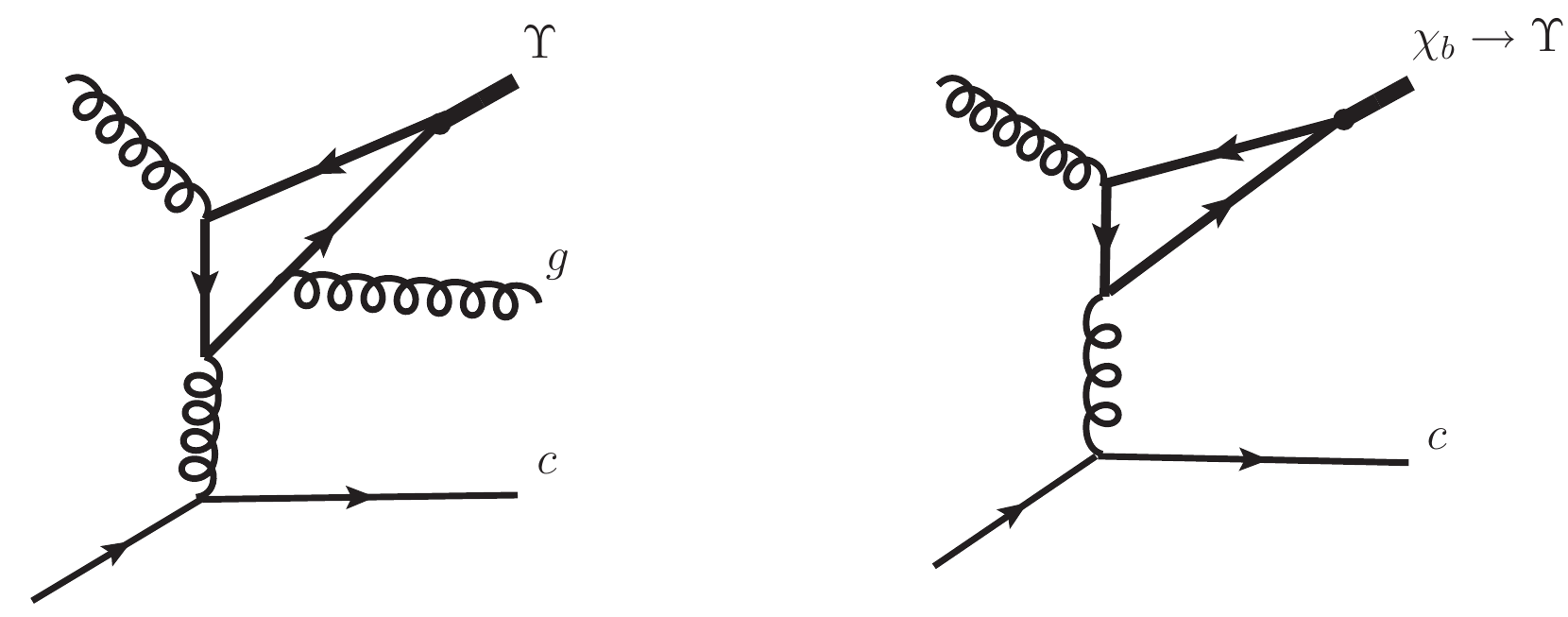}}
\caption{ The examples of diagrams? which contribute to the $\Upsilon +c$ production in the interaction with sea charm quark.}
\label{fig:c_sea_ups_cc_diagr}
\end{figure}

It was first shown in~\cite{Baranov:1997sg}, that the interaction with heavy sea quark can essentially contribute to the  multiple heavy quark production. But in our case the sea charm quark does not contribute essentially to $\Upsilon+c$ production.   The subprocess
\begin{equation}
gc \to \Upsilon_{\mathrm{direct}}+ g +c
\label{eq:gc_ups} 
\end{equation}
is suppressed by the additional order of $\alpha_s$ (see fig.~\ref{fig:c_sea_ups_cc_diagr}),
and the contribution of the subprocess
\begin{equation}
gc \to \chi_b(\to \Upsilon) +g +c
\end{equation}
by the order of magnitude is comparable with the contribution of (\ref{eq:gg_to_chib_cc}), and therefore,   it also can be neglected within  our rough analysis. Therefore, one can conclude that $c\bar c $-pair associated with  $\Upsilon$ in most cases is produced in gluon splitting. 

According our estimations within LO NRQCD for LHCb kinematics

 \begin{equation}
\frac{\sigma_{\mathrm{SPS}}^{\Upsilon + c}}{\sigma_{\mathrm{LHCb}}^{\Upsilon}} \sim  0.2\div 0.6 \%.
\end{equation}

It is worth to note, that there is an alternative way to estimate SPS. We could try to use the experimental probability value of the gluon splitting to $c\bar c$~pair.  According to the LEP data this probability $P^{g \to c\bar c}_\mathrm{LEP}$ is about 2.4\%~\cite{Akers:1994jc,Akers:1995wu}. 
Thus it could be supposed that gluon associated  with $\Upsilon$ quarkonium will produce $c$~quark in 2\% of events:
\begin{equation}
\frac{\sigma_{\mathrm{SPS}}^{\Upsilon + c}}{\sigma_{\mathrm{LHCb}}^{\Upsilon}} \approx     P^{g \to c\bar c}_\mathrm{LEP}      
\cdot k \sim 2\%,
\end{equation}
where $k$ is a geometrical  acceptance  of the LHCb detector, which can be approximately estimated as
\begin{equation}
k=\frac{ [\sigma^\mathrm{LO}(gg \to  \Upsilon_{\mathrm{direct}} + c \bar c)]_\mathrm{2.0<y_\mathrm{charm}<4.5}}{[\sigma^\mathrm{LO}(gg \to  \Upsilon_{\mathrm{direct}} + c \bar c)]_\mathrm{without~cuts~on~charm}}\approx 0.7.
\end{equation}

It is useful to mention, that the theoretical predictions of  $P^{g \to c\bar c}$ obtained  within the leading order calculation  ( $P^{g \to c\bar c}_\textrm{LO}=0.607\%$~\cite{Mangano:1992qq}), as well as within the resummed leading order calculation ($P^{g \to c\bar c}_\textrm{RSLO}=1.35\%$~\cite{Seymour:1994bz}) underestimate the LEP data.

\section{ DPS and accounting of charm quarks from PDF}

The yield of $\Upsilon$ mesons associated with open charm in DPS can be roughly estimated within formula~(\ref{eq:DPS}):
\begin{equation}
\frac{\sigma_{\mathrm{DPS}}^{\Upsilon + c}}{\sigma_{\Upsilon}}
=\frac{\sigma^c_\mathrm{LHCb}}{\sigma_\mathrm{eff}}\sim 10\%.
\end{equation}

Therefore DPS approach predicts ten times lager yield of of $\Upsilon$ mesons associated with open charm than LO SPS.

Also there is another method to estimate the cross section value of  $\Upsilon$ and open charm production. We could try to evaluate  the number of charm quarks, which "exists" in proton at the scale of order of the $\Upsilon$ mass, as follows: \begin{equation}
n_\mathrm{charm}\sim \int\limits_{x_\mathrm{min}}^{x_\mathrm{max}}f_\mathrm{charm}(x,Q)\,\mathrm{d}x,
\end{equation}
where $x_\mathrm{min}$ and $x_\mathrm{max}$ are determined by the the LHCb fiducial region.

\begin{figure}
 \centering
\resizebox*{0.7\textwidth}{!}{\includegraphics{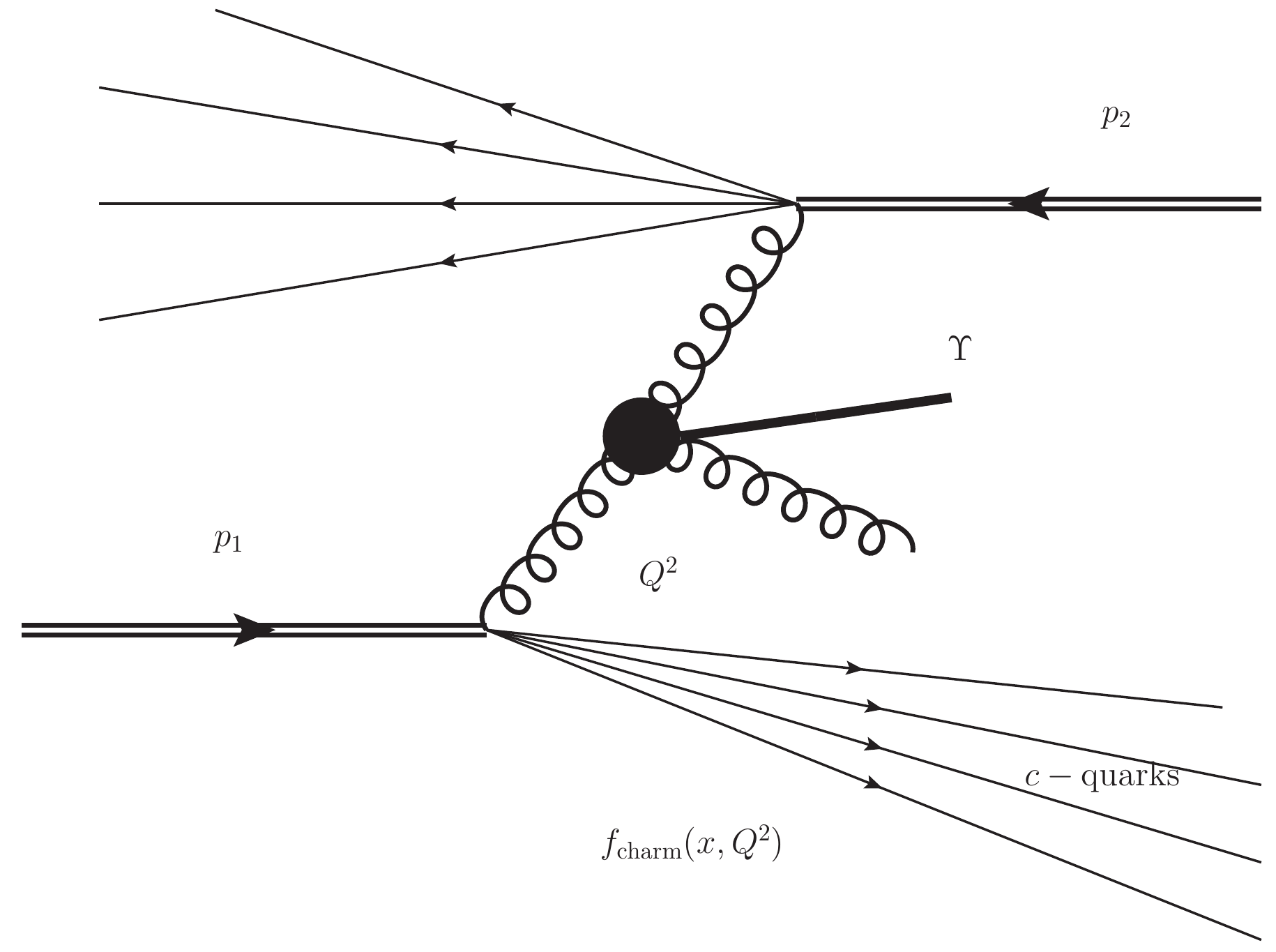}}
\caption{ Accounting of $c$-quarks from PDF at the $\Upsilon$ production scale.}
\end{figure}

Taking into account that $x \sim \frac{E_T}{\sqrt{s}}\exp(y) $ and assuming that $\langle E_T \rangle \sim 2.5$~GeV and $Q\sim 10$~GeV,  one can obtain the following very rough estimation for the LHCb kinematical cuts on charm hadrons ($2.0<y< 4.5$):
\begin{equation}
\frac{\sigma_{\Upsilon+c}}{\sigma_{\Upsilon}} \sim \int\limits_{0.0026}^{0.032}f_\mathrm{charm}(x,10~\mathrm{GeV})\,\mathrm{d}x\sim 50\%.
\end{equation}

This means that at scale of $\Upsilon$ mass in half of the cases a proton  "contains"  a charm quark.  Therefore one could suppose that  this charm quark transforms in the charm hadron during the nonperturbative destruction of the proton. 

\section{Conclusions}

We have shown  that the predictions of  SPS and DPS approaches for the associated production of $\Upsilon$ and a charmed hadron differ from each other by the order of magnitude.  
According SPS+LO the yield of $\Upsilon +c$ is about of $0.2\div 2\%$ of $\Upsilon$ production for the LHCb kinematics, whereas DPS or charm quark accounting in pdf predicts about 10\%. 

We think that the LHC experiments will obtain the $\Upsilon +c$ yield, which is close to the predicted in the framework  of DPS. This assurance is based on the fact, that the experimental data on $J/\psi + c$ and double open charm production are in fair agreement with    the  DPS predictions. 

It is known, that  the distributions on $p_T$ for the single $J/\psi$   production and for  $J/\psi$ associated with open charm are different. This difference can not be explained within the simplest version of DPS.  Maybe  it  the mutual hadronization of two $c \bar c$~pairs could influence the distribution shape. The $\Upsilon +c$ production is more pure case, and $p_T$ distribution of $\Upsilon$ in the single production and in the production associated with open charm should be more close to each other.

We thank I. Belyaev for the fruitful discussion. 
A. Berezhnoy acknowledges the support from MinES of RF (grant 14.610.21.0002, identification number RFMEFI61014X0002). The work of A. Likhoded is partially supported by  Russian Foundation for Basic Research (grant 15-02-03244 A).

\bibliography{/home/aber/papers/heavy_quark}
\end{document}